\documentclass[prl]{revtex4}
\begin{document}
\title{{\bf The General Boson Normal Ordering Problem}}
\author{\large{P. Blasiak$^{\dag}$,\footnotetext {$^{\dag}$e-mail: blasiak@lptl.jussieu.fr}
 K.A. Penson$^{\ddag}$\footnotetext{$^{\ddag}$e-mail:
penson@lptl.jussieu.fr} and A.I. Solomon$^{\S}$
\footnotetext{$^{\S}$e-mail: a.i.solomon@open.ac.uk}}\linebreak}
\affiliation{$^{{\dag}{\ddag}{\S}}$Universit\'{e} Pierre et Marie Curie, Laboratoire   de  Physique   Th\'{e}orique  des  Liquides,\linebreak
Tour 16, $5^{i\grave{e}me}$ \'{e}tage, 4, place Jussieu, F 75252 Paris
Cedex 05, France \linebreak\linebreak
$^{\dag}$H. Niewodnicza{\'n}ski Institute of Nuclear Physics,\linebreak
ul.Eliasza Radzikowskiego 152, PL 31-342 Krak{\'o}w, Poland}
\begin{abstract}
We solve the boson normal ordering problem for $F\left[(a^{\dag})^r
a^s\right]$, with $r,s$ positive integers,
$\left[a,a^{\dag}\right]=1$, i.e. we provide exact and explicit
expressions for its normal form ${\mathcal{N}}
\left\{F\left[(a^{\dag})^r a^s\right]\right\}$, where in ${\mathcal N}
\left(F\right)$ all $a$'s are to the right. The solution involves
integer sequences of numbers which are
generalizations  of the conventional Bell and Stirling numbers
whose values they assume  for $r=s=1$. A comprehensive theory of such
generalized combinatorial numbers is given including closed-form
expressions (extended Dobinski - type formulas) and generating
functions.
These last are special
expectation values in boson coherent states.
\end{abstract}
            \maketitle
\bigskip

Consider a function $F(x)$ having a Taylor expansion around
$x=0$, i.e. $F(x)=\sum_{k=0}^{\infty} \frac{F^{(k)}(0)}{k!}x^k$.
In this note we will collect the formulas concerning our solution
of the normal ordering problem for $F\left[(a^{\dag})^r
a^s\right]$, where $a, a^{\dag}$ are the boson annihilation and
creation operators, $\left[a,a^{\dag}\right]=1$, and $r$ and $s$
are positive integers. The normally ordered form of the operator
$F\left[(a^{\dag})^r a^s\right]$ is denoted by ${\mathcal{N}}
\left\{F\left[(a^{\dag})^r a^s\right]\right\}$, where in
${\mathcal N}(F)$ all the $a$'s are to the right. It satisfies
the operator identity:
\begin{eqnarray}
   {\mathcal{N}} \left\{F\left[(a^{\dag})^r a^s\right]\right\}=
   F\left[(a^{\dag})^r a^s\right].\label{Q}
\end{eqnarray}
Furthermore, an auxiliary symbol $:O(a,a^{\dag}):$ will be used,
which means expand $O$ in powers of $a$ and $a^{\dag}$ and
order normally assuming they commute \cite{KlaudSud},\cite{Luisell}.

The combinatorial numbers $S(n,k)$, known as Stirling numbers of the
second kind, and their sums $B(n)$, the Bell numbers, arise naturally
in the normal ordering procedure for $r=s=1$, as follows \cite{KatrielNC}:
\begin{eqnarray}
   (a^{\dag} a)^n=\sum_{k=1}^nS(n,k)(a^{\dag})^ka^k,
\end{eqnarray}
\begin{eqnarray}
B(n)=\sum_{k=1}^nS(n,k),
\end{eqnarray}
which may be taken as definitions of $S(n,k)$ and $B(n)$. In the
present work we shall treat the case of general $(r,s)$, consequently
generalizing these combinatorial numbers.

For the moment we restrict ourselves to the case $r\geq s$, the
other alternative being treated later. We do not give the proofs
of the formulas here; they will be given elsewhere \cite{BPS}. The
case $r=s=1$ is known \cite{KlaudSud},\cite{Luisell} and some
features of the $r>1, s=1$ case have been published \cite{Lang}.
When needed we shall refer to \cite{KlaudSud},\cite{Luisell} and
\cite{Lang} where particular cases of our more general formulas
are discussed.

We define the set of positive integers $S_{r,s}(n,k)$ entering the expansion:
\begin{eqnarray}
    \left[(a^{\dag})^r a^s\right]^n=
    {\mathcal{N}} \left\{\left[(a^{\dag})^r a^s\right]^n\right\}=
    (a^{\dag})^{n(r-s)}\left[\sum_{k=s}^{ns}S_{r,s}(n,k)(a^{\dag})^ka^k\right],\label{P}
\end{eqnarray}
for $n=1,2,...$ . Once the $S_{r,s}(n,k)$ are known, the normal ordering of
$\left[(a^{\dag})^r a^s\right]^n$ is achieved. The same applies to any
operator $F\left[(a^{\dag})^r a^s\right]$ if $F(x)$ has a Taylor expansion
around $x=0$. The row sums of the triangle
$S_{r,s}(n,k)$ are given by:
\begin{eqnarray}
    B_{r,s}(n)=\sum_{k=s}^{ns} S_{r,s}(n,k),\label{Z}
\end{eqnarray}
which for any $t$ extend to the polynomials of order $ns$ defined by:
\begin{eqnarray}\label{A}
    B_{r,s}(n,t)=\sum_{k=s}^{ns} S_{r,s}(n,k)t^k.
\end{eqnarray}
All the subsequent formulas are consequences of the following
relation, linking the polynomial of Eq.(\ref{A}) to a certain
infinite series:
\begin{eqnarray}\label{B}
    e^{-t}\sum_{k=s}^{\infty}\frac{1}{k!}
    \prod_{j=1}^n \left[\left(k+(j-1)(r-s)\right)
    \cdot\left(k+(j-1)(r-s)-1\right)\right.\cdot\nonumber\\
    \left. \ldots\cdot\left(k+(j-1)(r-s)-s+1\right)\right]t^k= \\
    =\sum_{k=s}^{ns}S_{r,s}(n,k)t^k,\ \ \ \ \ \ \  \ \ \ \ \ \  n=1,2...\nonumber
\end{eqnarray}
which  for $r,s\geq 1$ and $t=1$ is an analogue of the celebrated
Dobinski relation, which expresses the combinatorial Bell numbers
$B_{1,1}(n)$ as a sum of an infinite series
\cite{Wilf},\cite{Const},\cite{Comtet}:
\begin{eqnarray}
    B_{1,1}(n)=\frac{1}{e}\sum_{k=0}^{\infty}\frac{k^n}{k!}.
\end{eqnarray}
For a review of various characteristics of Bell numbers
$B_{1,1}(n)$ see \cite{Aigner}. It seems  that Bell and Stirling
numbers are now beginning to appear in textbooks of mathematical
physics \cite{Aldrov}.

By setting $t=1$ in Eq.(\ref{B}) we obtain expressions for the
{\em generalized Bell numbers} $B_{r,s}(n),n=1,2...$ :
\begin{eqnarray}
    r>s:\ \ &B_{r,s}(n)=B_{r,s}(n,1)&=\frac{1}{e}\sum_{k=s}^\infty\frac{1}{k!}\prod_{j=1}^n\left[k+(j-1)(r-s)\right]^{\underline{s}} \label{C}\\
    &&=\frac{(r-s)^{s(n-1)}}{e}\sum_{k=0}^\infty\frac{1}{k!}\left[\prod_{j=1}^s
    \frac{\Gamma(n+\frac{k+j}{r-s})}{\Gamma(1+\frac{k+j}{r-s})}\right],\label{D}\\
    &&\nonumber\\
    r=s:\ \ &B_{r,r}(n)=B_{r,r}(n,1)&=\frac{1}{e}\sum_{k=0}^\infty \frac{1}{k!}\left[\frac{(k+r)!}{k!}\right]^{n-1}.\label{I}
\end{eqnarray}
For all $r,s$ we set $B_{r,s}(0)=1$ by convention.
In Eq.(\ref{C}) we use the notation $m^{\underline{s}}:=m(m-1)(m-2)\ldots(m-s+1)$
for the falling factorial \cite{Knuth}, and in Eq.(\ref{D}) $\Gamma(y)$ is the Euler
gamma function.

The numbers $S_{r,s}(n,k)$ are non-zero for $s\leq k\leq ns$, and
satisfy by convention $S_{r,s}(n,0)=\delta_{n,0}$. We shall refer
to them as  {\em  generalized Stirling numbers of the second
kind}. Their exact expressions in the form of a finite sum are:
\begin{eqnarray}
    S_{r,s}(n,k)&=&\frac{(-1)^k}{k!}\sum_{p=s}^k(-1)^p{\left(\begin{array}{c}k\\p\end{array}\right)}
    \prod_{j=1}^n\left(p+(j-1)(r-s)\right)^{\underline{s}}\label{E}\\
    &=&\frac{(-1)^k}{k!}\left\{\left(x^r\frac{d^s}{dx^s}\right)^n
    \left[(1-x)^k-\sum_{p=0}^{s-1}{\left(\begin{array}{c}k\\p\end{array}\right)}(-x)^p\right]\right\}_{x=1},\label{E2}
\end{eqnarray}
from which for $r=s=1$ one obtains the standard form of the classical
Stirling numbers of the second kind \cite{Comtet}:
\begin{eqnarray}\label{F}
    S_{1,1}(n,k)=\frac{(-1)^k}{k!}\sum_{p=1}^{k}(-1)^p\left(\begin{array}{c}k\\p\end{array}\right)p^n,
\end{eqnarray}
and for $r=2, s=1$,
\begin{eqnarray}
    S_{2,1}(n,k)=\frac{n!}{k!}\left(\begin{array}{c}n-1\\k-1\end{array}\right),
\end{eqnarray}
which are the so-called unsigned Lah numbers
\cite{Lang},\cite{Comtet}.

That the numbers $S_{r,s}(n,k)$ are natural extensions of
$S_{1,1}(n,k)$ can be neatly seen by observing their action on the
space of polynomials generated by falling factorials. Let $x$ be
any real number. Then,
\begin{eqnarray}
    \prod_{j=1}^n(x+(j-1)(r-s))^{\underline{s}}=\sum_{k=s}^{ns}S_{r,s}(n,k)x^{\underline{k}}.
\end{eqnarray}
This last equation when specified to $r=s$ gives a particularly transparent
interpretation of $S_{r,r}(n,k)$ as connection coefficients:
\begin{eqnarray}
    (x^{\underline{r}})^n=\sum_{k=r}^{nr}S_{r,r}(n,k)x^{\underline{k}}.
\end{eqnarray}
For $r=s=1$ it boils down to the defining equations for
$S_{1,1}(n,k)$ in terms of $x^{\underline{k}}$, used by J. Stirling himself
\cite{Knuth}:
\begin{eqnarray}
    x^n=\sum_{k=1}^nS_{1,1}(n,k)x^{\underline{k}}.
\end{eqnarray}

The relations for $S_{r,s}(n,k)$ become particularly appealing for
$r=s$: Eq.(\ref{E}) can be reformulated as:
\begin{eqnarray}
    S_{r,r}(n,k)
    &=&\frac{(-1)^k}{k!}\sum_{p=r}^{k}(-1)^p\left(\begin{array}{c}k\\p\end{array}\right)
    \left(p^{\underline{r}}\right)^n,\label{K}
\end{eqnarray}
which differs from Eq.(\ref{F}) in that the sum in Eq.(\ref{K})
starts from $p=r$ and $p^n$ becomes $(p^{\underline{r}})^n$. The
recurence relations for $S_{r,r}(n,k)$ are the following:
\begin{eqnarray}
    &&S_{r,r}(1,r)=1,\quad
    S_{r,r}(n,k)=0,\ \ \qquad\qquad\qquad\qquad\qquad\qquad\qquad k<r,\quad nr<k\leq (n+1)r,\\
    &&S_{r,r}(n+1,k)=\sum_{p=0}^r\left(\begin{array}{c}k+p-r\\p\end{array}\right)
    r^{\underline{p}}\ S_{r,r}(n,k+p-r),\qquad\qquad r\leq k\leq nr,\quad n>1,
\end{eqnarray}
where the definition $r^{\underline{0}}:=1$ is made.
Note that for $r=s=1$ we get the known relation for conventional Stirling numbers
i.e.: $S_{1,1}(n+1,k)=kS_{1,1}(n,k)+S_{1,1}(n,k-1)$ with appropriate
initial conditions \cite{Knuth}.

The ``non-diagonal'' generalized Bell numbers $B_{r,s}(n)$ can
always be expressed as special values of generalized
hypergeometric functions $_pF_q$. Algebraic manipulation of
Eqs.(\ref{C})-(\ref{I}) yields the following examples.
For $r>1, s=1$,
$B_{r,1}(n)$ is a combination of $r-1$ different hypergeometric
functions of type $_1F_{r-1}(\ldots;x)$, each of them evaluated at the same
value of argument $x=(r-1)^{1-r}$; here are some lowest order cases:
\begin{eqnarray}
    B_{2,1}(n)&=&\frac{n!}{e}{_1F_1}(n+1;2;1)=(n-1)!L_{n-1}^{(1)}(-1),\label{J}\\
    B_{3,1}(n)&=&\frac{2^{n-1}}{e}\left(\frac{2\Gamma(n+\frac{1}{2})}{\sqrt{\pi}}
    {_1F_2}\left(n+\frac{1}{2};\frac{1}{2},\frac{3}{2};\frac{1}{4}\right)+
    n!{_1F_2}\left(n+1;\frac{3}{2},2;\frac{1}{4}\right)\right),
\end{eqnarray}
etc. In Eq.(\ref{J}) $L_m^{(\alpha)}(y)$ is the associated Laguerre polynomial.

Similarly the series $B_{2r,r}(n)$ can be written down in
a compact form using the confluent hypergeometric function of Kummer:
\begin{eqnarray}
    B_{2r,r}(n)=\frac{(rn)!}{e\cdot r!}{_1F_1}(rn+1,r+1;1).\label{Y}
\end{eqnarray}

In contrast, the "diagonal" numbers $B_{r,r}(n)$ of Eq.(\ref{I}), which also can be
rewritten as:
\begin{eqnarray}
    B_{r,r}(n)=\frac{1}{e}\sum_{k=0}^\infty
\frac{1}{(k+r-1)!}\left[k(k+1)\ldots(k+r-1)\right]^n,\ \ \ \  n=1,2,\ldots
\end{eqnarray}
cannot be expressed through hypergeometric functions. However, the
sequences $B_{r,r}(n)$ possess a particularity that they can always be
expressed in terms of conventional Bell numbers and $r$-nomial
(binomial, trinomial\ldots) coefficients. For example:
\begin{eqnarray}
    B_{2,2}(n)=\sum_{k=0}^{n-1}\left(\begin{array}{c}n-1\\k
    \end{array}\right) B_{1,1}(n+k).
\end{eqnarray}

Some low-order triangles of $S_{r,s}(n,k)$ and their associated $B_{r,s}(n)$
are presented in Table 1.

It turns out that various generating functions for $B_{r,s}(n)$ and
$S_{r,s}(n,k)$ can be related to special quantum states, which are called
coherent states \cite{KlaudSud}, defined as linear combinations of the eigenstates of
the number operator, $N=a^{\dag}a$, $N|n\rangle =n|n\rangle$, $\langle n|n'\rangle =\delta_{n,n'}$ and
defined as:
\begin{eqnarray}
    |z\rangle=e^{-\frac{|z|^2}{2}}\sum_{n=0}^\infty
    \frac{z^n}{\sqrt{n!}}|n\rangle,
\end{eqnarray}
(with $\langle z|z\rangle =1$), for $z$ complex. The states $|z\rangle$ satisfy:
\begin{eqnarray}
    a|z\rangle =z|z\rangle .
\end{eqnarray}
It has been noticed in \cite{Katriel} that,
\begin{eqnarray}
    \langle z|(a^{\dag} a)^n|z\rangle \stackrel{|z|=1}{=}B_{1,1}(n)
\end{eqnarray}
and the exponential generating function (egf) of the numbers
$B_{1,1}(n)$, i.e. $\sum_{n=0}^\infty B_{1,1}(n)\frac{\lambda^n}{n!}$ satisfies
\begin{eqnarray}
    \langle z|e^{\lambda a^{\dag} a}|z\rangle =
    \langle z|:e^{a^{\dag}a(e^\lambda-1)}:|z\rangle \stackrel{|z|=1}{=}e^{e^\lambda-1}=\sum_{n=0}^\infty B_{1,1}(n)\frac{\lambda^n}{n!}
\end{eqnarray}
which is a restatement of the known fact that \cite{KlaudSud},\cite{Luisell},\cite{Lang}:
\begin{eqnarray}
   e^{\lambda a^{\dag} a}={\mathcal{N}}\left(e^{\lambda a^{\dag} a}\right)=:e^{a^{\dag} a (e^\lambda-1)}:\ .
\end{eqnarray}
How does it generalize to $r,s\geq1$ ?

Eqs.(\ref{P}) and (\ref{Z}) give directly:
\begin{eqnarray}\label{T}
    \langle z|\left[(a^{\dag})^ra^s\right]^n|z
\rangle \stackrel{z=1}{=}B_{r,s}(n),
\end{eqnarray}
valid for all $r,s$.

We first treat  the case $r=s=1,2\ldots$. Note in this context that a Hermitian
Hamiltonian $(a^{\dag})^ra^r$ is of great importance in quantum
optics, as for $r=2,3,\ldots$ it provides a description of
non-linear Kerr-type media\cite{MandelWolf}. Using Eq.(\ref{K})
we can find the following egf of $S_{r,r}(n,k)$:
\begin{eqnarray}
    \sum_{n=\lceil k/r\rceil }\frac{x^n}{n!}S_{r,r}(n,k)=
    \frac{(-1)^k}{k!}\sum_{p=r}^k(-1)^p\left(\begin{array}{c}k\\p\end{array}\right)
    \left(e^{xp(p-1)\ldots(p-r+1)}-1\right),
\end{eqnarray}
where $\lceil y\rceil$ is the {\em ceiling} function \cite{Knuth}, defined as
the nearest integer greater or equal to $y$. This yields via
exchange of order of summation:
\begin{eqnarray}
    e^{\lambda (a^{\dag})^ra^r}={\mathcal{N}}\left(e^{\lambda (a^{\dag})^ra^r}\right)=
    1+:\sum_{k=r}^\infty \frac{(-1)^k}{k!}\left[\sum_{p=r}^k(-1)^p\left(\begin{array}{c}k\\p\end{array}\right)
    \left(e^{\lambda p(p-1)\ldots(p-r+1)}-1\right)\right](a^{\dag} a)^k:
\end{eqnarray}
and
\begin{eqnarray}
    \langle z|e^{\lambda (a^{\dag})^r a^r}|z\rangle \stackrel{|z|=1}{=}
    1+\sum_{k=r}^\infty \frac{(-1)^k}{k!}\left[\sum_{p=r}^k(-1)^p\left(\begin{array}{c}k\\p\end{array}\right)
    \left(e^{\lambda p^{\underline{r}}}-1\right)\right].\label{L}
\end{eqnarray}
Another case which can be written in a closed form is $r>1$, $s=1$,
for which the egf for $S_{r,1}(n,k)$ reads:
\begin{eqnarray}
    \sum_{n=\lceil k/r\rceil }^\infty \frac{x^n}{n!}S_{r,1}(n,k)=\frac{1}{k!}
    \left\{\left(1-(r-1)x\right)^{-\frac{1}{r-1}}-1 \right\}.
\end{eqnarray}
One then obtains  \cite{Lang},\cite{Dattoli}:
\begin{eqnarray}
    e^{\lambda (a^{\dag})^ra}={\mathcal{N}}\left(e^{\lambda (a^{\dag})^ra}\right)=
    :exp\left\{\left[(1-\lambda(a^{\dag})^{r-1}(r-1))^{-\frac{1}{r-1}}-1\right]a^{\dag} a \right\}:\label{M}
\end{eqnarray}
For arbitrary $r>s$ the following formula is still valid:
\begin{eqnarray}
    e^{\lambda (a^{\dag})^ra^s}={\mathcal{N}}\left(e^{\lambda (a^{\dag})^ra^s}\right)=
    1+:\sum_{n=1}^\infty
\frac{\lambda^n}{n!}(a^{\dag})^{n(r-s)}\left(\sum_{k=s}^{ns}S_{r,s}(n,k)(a^{\dag} a)^k\right):\label{O}
\end{eqnarray}
where the explicit form of $S_{r,s}(n,k)$ may be used, see
Eqs.(\ref{E})-(\ref{E2}).

A close look at Eqs.(\ref{L}) and (\ref{T}) reveals that
\begin{eqnarray}
    \langle z|e^{\lambda (a^{\dag})^r a^r}|z\rangle \stackrel{|z|=1}{=}\sum_{n=0}^\infty B_{r,r}(n)\frac{\lambda^n}{n!}.\label{N}
\end{eqnarray}
Comparing Eqs.(\ref{N}),(\ref{M}) and (\ref{L}) we conclude that
for $r>1, s=1$ and $r=s$ the egf of respective generalized Bell
numbers are special matrix elements of $e^{\lambda(a^{\dag})^ra^s}$ in coherent states.

However, for arbitrary $r>s>1$ examination of
Eqs.(\ref{C})-(\ref{D}) confirms that the numbers $B_{r,s}(n)$
increase so rapidly with $n$ that one cannot define their egf's
meaningfully. In  particular, this is reflected by  Eq.(\ref{O})
which is true in its operator form, but one needs to exercise
care  when calculating its matrix elements  as the convergence of
the result may require limitations on $\lambda$.

A well-defined and convergent procedure for such sequences is to consider what we call
\emph{hypergeometric generating functions} (hgf), i.e. the egf for the
ratios $B_{r,s}/(n!)^t$, where $t$ is an appropriately chosen
integer. A case in point is the series $B_{3,2}(n)$ which may be written
explicitly from  Eq.(\ref{D}) as:
\begin{eqnarray}
    B_{3,2}(n)=\frac{1}{e}\sum_{k=0}^\infty
    \frac{(n+k)!(n+k+1)!}{k!(k+1)!(k+2)!}.
\end{eqnarray}
Its hgf $G_{3,2}(\lambda)$ is then:
\begin{eqnarray}
    G_{3,2}(\lambda)=\sum_{n=0}^\infty\left[\frac{B_{3,2}(n)}{n!}\right]\frac{\lambda^n}{n!}
    =\frac{1}{e}\sum_{k=0}^\infty\frac{1}{(k+2)!}{_2F_1(k+2,k+1;1;\lambda)}.
\end{eqnarray}
Similarly one can work out other examples of hgf. See \cite{SPS} for
other instances where this type of hgf appear.

All the previous considerations also apply  to the case $r<s$.  In this case we define  the generalized Stirling numbers by (note the difference with Eq.(\ref{P})):
\begin{eqnarray}
    r\leq s:\qquad\qquad\left[(a^{\dag})^r a^s\right]^n=
    {\mathcal{N}} \left\{\left[(a^{\dag})^r a^s\right]^n\right\}=
    \left[\sum_{k=r}^{nr}S_{r,s}(n,k)(a^{\dag})^ka^k\right]a^{n(s-r)},\label{U}
\end{eqnarray}

By taking the Hermitian conjugate of Eq.(\ref{P}), with the change $r\leftrightarrow s$, we obtain
\begin{eqnarray}
   r\leq s:\qquad\qquad \left[(a^{\dag})^r a^s\right]^n=
    \left[\sum_{k=r}^{nr}S_{s,r}(n,k)(a^{\dag})^ka^k\right]a^{n(s-r)},
\end{eqnarray}\label{V}
whence the symmetry relation:
\begin{eqnarray}
    &&S_{r,s}(n,k)= S_{s,r}(n,k),\ \ \ \ \qquad\qquad\qquad
    s\leq k\leq ns,\ \ (r\leq s).
\end{eqnarray}

The normal ordering of $F[a^s(a^{\dag})^r]$ involves the {\em antinormal to
normal} order, in the terminology of \cite{KatDuch}. With this in mind
we define the \emph{anti}-Stirling numbers of the second kind
$\tilde S_{r,s}(n,k)$ as:
\begin{eqnarray}
    \quad\  r\geq s:\qquad\qquad [a^s(a^{\dag})^r]^n=(a^{\dag})^{n(r-s)}\left[\sum_{k=0}^{ns}\tilde S_{s,r}(n,k)(a^{\dag})^ka^k\right],
\end{eqnarray}
and similarly \emph{anti}-Bell numbers: $\tilde B_{r,s}(n)=\sum_{k=0}^{ns}
\tilde S_{r,s}(n,k)$ for $r\geq s$.

It can be demonstrated \cite{BPS} that they satisfy the following symmetry
properties:
\begin{eqnarray}
    &&\tilde S_{s,r}(n,k)=\tilde S_{r,s}(n,k),\ \ \ \
\qquad\qquad\qquad 0\leq k\leq ns,\ \ (r\geq s),\\
    &&\tilde S_{r,s}(n,k)= S_{r,s}(n+1,k+s),\qquad\qquad 0\leq k\leq ns,\ \
(r\geq s),
\end{eqnarray}
and $\tilde B_{r,s}(n)= B_{r,s}(n+1)$.

If in Eq.(\ref{Q}) $F(x)$ has a Taylor expansion of $F(x)$ around $x_0\neq0$, then
\begin{eqnarray}
    F\left[(a^{\dag})^ra^s\right]={\mathcal{N}}\left\{F\left[(a^{\dag})^ra^s\right]\right\}=
    \sum_{k=0}^\infty
    \frac{F^{(k)}(x_o)}{k!}{\mathcal{N}}\left\{\left[(a^{\dag})^ra^s-x_o\right]^k\right\},
\end{eqnarray}
and our above results should be supplemented by the appropriate binomial
expansion coefficients.

All the $B_{r,s}(n)$ described here are the $n$-th moments of positive functions on the positive half axis; some solutions of the associated Stieltjes moment problem have been recently discussed at some length \cite{PS}. In addition we have found that even larger classes of
combinatorial sequences are solutions of the moment problem and they
can be used for the construction of new families of coherent states \cite{PSKrak}.

We are considering the  possible combinatorial interpretation of the above results.
\\\linebreak
\linebreak
\\
{\bf Acknowledgments}
\\

We thank G. Duchamp, L. Haddad, A. Horzela, A. Lascoux, M. Mendez and J.Y. Thibon for
enlightening discussions. N.J.A. Sloane's Encyclopedia of
Integer Sequences (http://www.research.att.com/{\textasciitilde}njas/sequences) was an
essential help in sorting out the properties of sequences we have
encountered in this work, as was \emph{Maple} in the evaluation of the
algebraic expressions.

\begin{eqnarray}
 \underline{\bf r=1, s=1}\ \ \ \ \qquad\qquad\qquad\qquad\qquad\qquad&\qquad\qquad\qquad& \underline{\bf r=2, s=1}\nonumber\\
\nonumber\\
S_{1,1}(n,k),\ 1\leq k\leq n\qquad B_{1,1}(n)&&\ \qquad\qquad S_{2,1}(n,k),\ 1\leq k\leq n\qquad\qquad\qquad B_{2,1}(n)\nonumber\\
    \begin{array}{cl|lllllllllllcc}\cline{3-14}&&&&&&\\
    n=1&&&1&&&&&&&&&&1&\\
        n=2&&&1&1&&&&&&&&&2&\\
        n=3&&&1&3&1&&&&&&&&5&\\
        n=4&&&1&7&6&1&&&&&&&15&\\
        n=5&&&1&15&25&10&1&&&&&&52&\\
    n=6&&&1&31&90&65&15&1&&&&&203&
    \end{array}&\ \ \ & \begin{array}{cl|lllllllllllccc}\cline{3-14}&&&&&&\\
    n=1&&&1&&&&&&&&&&1&\\
        n=2&&&2&1&&&&&&&&&3&\\
        n=3&&&6&6&1&&&&&&&&13&\\
        n=4&&&24&36&12&1&&&&&&&73&\\
        n=5&&&120&240&120&20&1&&&&&&501&\\
    n=6&&&720&1800&1200&300&30&1&&&&&4051&
    \end{array}\nonumber
\end{eqnarray}
\newline
\newline

\underline{\bf r=2, s=2}
\begin{eqnarray}
    S_{2,2}(n,k),\ 2\leq k\leq 2n\ \quad\qquad\qquad\qquad\qquad\qquad\qquad\quad\qquad\qquad\qquad\qquad\qquad B_{2,2}(n)\ &&\nonumber\\
    \begin{array}{cl|lllllllllllcccccccc}\cline{3-19}&&&&&&\\
    n=1&&&1&&&&&&&&&&&&&&&1\\
        n=2&&&2&4&1&&&&&&&&&&&&&7\\
        n=3&&&4&32&38&12&1&&&&&&&&&&&87\\
        n=4&&&8&208&652&576&188&24&1&&&&&&&&&1657\\
        n=5&&&16&1280&9080&16944&12052&3840&580&40&1&&&&&&&43833\\
    n=6&&&32&7744&116656&412800&540080&322848&98292&16000&1390&60&1&&&&&1515903\\
    \end{array}&&\nonumber
\end{eqnarray}
\newline
\newline

\underline{\bf r=3, s=2}
\begin{eqnarray}
    S_{3,2}(n,k),\ 2\leq k\leq 2n\ \quad \qquad\quad\qquad\qquad\qquad\qquad\qquad\qquad\qquad\qquad\qquad\quad\qquad\qquad\qquad\qquad\qquad B_{3,2}(n)\ \ &&\nonumber\\
    \begin{array}{cl|lllllllllllcccccccc}\cline{3-19}&&&&&&\\
    n=1&&&1&&&&&&&&&&&&&&&1\\
        n=2&&&6&6&1&&&&&&&&&&&&&13\\
        n=3&&&72&168&96&18&1&&&&&&&&&&&355\\
        n=4&&&1440&5760&6120&2520&456&36&1&&&&&&&&&16333\\
        n=5&&&43200&259200&424800&285120&92520&15600&1380&60&1&&&&&&&1121881\\
    n=6&&&1814400&15120000&34776000&33566400&16304400&4379760&682200&62400&3270&90&1&&&&&106708921\\
    \end{array}&&\nonumber
\end{eqnarray}
\newline

\end{small}

\end{document}